\documentclass[12pt]{article}

\usepackage{bm}
\usepackage{amssymb}
\usepackage{enumerate}

\usepackage{graphicx} 
\begin{document}

\title{Bounds on the Gell-Mann ~-~Low functions  in quantum electrodynamics and in the Wess-Zumino model} 
\author{N.V.Krasnikov
\\INR RAS,   Moscow 117312, Russia
\\and
\\ JINR, Dubna 141980, Russia
}
\maketitle
\begin{abstract}
  We derive bounds
$ |\frac{d\psi(\alpha)}{d\alpha}| \leq 1 $, $  \frac{d(\frac{d\psi(\alpha)}{d\alpha}\psi(\alpha))}{d\alpha} \leq 1  $
  on the GL (Gell-Mann ~-~Low) function $\psi(\alpha)$  from the Kallen-Lehmann dispersion representation
   in quantum electrodynamics.
 We also derive analogous bounds  for         the GL function in the Wess-Zumino model. 
    The implications of the obtained inequalities are discussed. In particular, we obtain bounds on coupling
    constants in dark photon   model and in  dark matter model  with vector $(B-L)$ messenger.

\end{abstract}
\newpage
\section{Introduction}
In quantum field theory renormalization group functions determine the evolution of the theory in
ultraviolet or infrared regions \cite{bog1}.
For Green function $G_n(p_1,...,p_n) = \int exp(ip_kx_k)  <0|T(O_1(x_1)...O_n(x_n)|0>d^4x_1...d^4x_n   $ with local operators $O_k(x)$
the renormalization group equation reads\footnote{Here we consider the
  model with  massless particles and the single coupling constant. Besides we
  assume that the matrix of the anomalous dimensions is diagonal.}
\begin{equation}
  (\mu^2\frac{d}{d\mu^2} + \beta(\alpha)\frac{d}{d\alpha} + \sum_{i} \gamma_i(\alpha))G_n =0 \,.
\label{form1a}
\end{equation}
At present state of art for realistic models we can calculate the $\beta$-functions and the
anomalous dimensions $\gamma_k(\alpha)$ only within the perturbation theory. Therefore it is very interesting to obtain some information
on the behaviour of the renormalization group functions beyond the perturbation theory. In refs.\cite{Kr1, Kr2, Yam} the inequality
\begin{equation}
  0 \leq \psi(\alpha) \leq \alpha  \,
\label{form1bb}
\end{equation}
for the GL function \cite{Pet, Gell} in QED (quantum electrodynamics).
was  derived from the KL (Kallen-Lehmann) dispersion relation \cite{KL1, KL2} for the photon  propagator.


In this paper   using the KL representation  we derive new  inequalities 
\begin{equation}
   |\frac{d\psi(\alpha)}{d\alpha}| \leq 1 \,,
  \label{GL1}
\end{equation}
\begin{equation}
  ( \frac{d\psi(\alpha)}{d\alpha}\psi(\alpha))' \leq 1  \,
  \label{GL2}
\end{equation}
for the GL function  $\psi(\alpha)$ in QED and in the Wess-Zumino model.
Here  $f(\alpha)' \equiv \frac{df(\alpha)}{d\alpha}$.
The  inequalities  (\ref{GL1}, \ref{GL2}) are more strong than the inequality (\ref{form1bb}).
We also derive analogous bounds  for   the GL function in the Wess-Zumino model. 
Possible implications of derived inequalities are discussed. In particular, we obtain bound on new particles contribution
for the extension of the $SU_c(3)\otimes SU_L(2)\otimes U(1)$ SM model with additional scalar and fermion fields. 
  Also we obtain  bounds on the coupling
constants in  dark matter models with dark photon and   vector $(B - L)$ messenger.

The organization of the paper is the following. In the next section we derive general inequalities for the renormalization group
functions on the example of QED. Also in this section we  derive bound on the  GL function in the Wess-Zumino model.
 In section 3 we    discuss possible implications of the obtained results.  
Section 4 contains concluding remarks.

\section{General inequalities}

The KL representation for the transverse part of the photon propagator in QED  has the form \cite{bog1}
\begin{equation}
  D^{tr}(k^2, \alpha_0, m) = \frac{1}{k^2 + i\epsilon } + \int^{\infty}_0 \frac{\rho(t, \alpha_0, m)}{k^2 -t + i\epsilon} dt \,,
    \label{form1}
\end{equation}
where
\begin{equation}
  \int \exp(iqx)<0|T(A_{\mu}(x) A_{\nu}(o)|0> d^4x = -i(g_{\mu\nu} - i \frac{k_{\mu}k_{\nu}}{k^2})D^{tr}(k^2, \alpha_0, m) +
 i \frac{k_{\mu}k_{\nu}}{(k^2)^2}d^l(k^2) \,
 \label{form1aa}
\end{equation}
and
\begin{equation}
  \rho(t, \alpha_0, m) \geq 0 \,.
  \label{form1b}
\end{equation}
Here $m$ is  the electron mass and $\alpha_0 = \frac{1}{137}$ is the fine structure
constant. We use the KL representation (\ref{form1}) without subtractions since in perturbation theory additional
subtractions are not necessary. We assume that the KL representation   (\ref{form1}) without subtractions
is valid irrespective of the perturbation theory.
Really, the existence of arbitrary subtraction constants in the KL representation for the transverse photon propagator means
that physics depends not only on the electron mass $m$ and coupling constant $\alpha_0  $ but also
from other unknown parameters.
    The invariant charge   \cite{bog1}   in QED is proportional to the transverse part of photon propagator
\begin{equation}
  \bar{\alpha}(x, y, \alpha) =
  \alpha_0k^2  D^{tr}(k^2, \alpha_0, m) \,,
\label{form2}
\end{equation}
where $x = \frac{-k^2}{\mu^2}  >0 $, $ y = \frac{m^2}{\mu^2} $ and
\begin{equation}
  \alpha =  \bar{\alpha}(x=1, y, \alpha) \,.
  \label{form3}
\end{equation}
Using the KL representation (\ref{form1}) and the definition (\ref{form2}) of the invariant charge one can find that
  \begin{equation}
    \bar{\alpha}(x,y,\alpha) = x\int^{\infty}_0\frac{\bar{\rho}(t, \alpha, y)}{x + t}   dt \,,
\label{form4}
  \end{equation}
  where $ \bar{\rho}(t, \alpha, y)   = \alpha_0(\delta(t) + \mu^2\rho(t\mu^2, \alpha_o, m))  \geq 0 $.
  The renormalization condition for the invariant charge $\bar{\alpha}(x,y,\alpha)$ is
  \begin{equation}
    \alpha = \bar{\alpha}(1, y,\alpha) =  \int^{\infty}_0\frac{\bar{\rho}(t, \alpha, y)}{1 + t} dt \,.
    \label{form5}
\end{equation}
The renormalization group equation for the invariant charge has the form \cite{bog1, Pet, Gell}
\begin{equation}
  x \frac{\partial\bar{\alpha}(x,y,\alpha)}{\partial x} = \psi(\frac{y}{x}, \bar{\alpha}) \,,
\label{form6}
\end{equation}
where
\begin{equation}
  \psi(y, \alpha) = F(x =1, y, \alpha) \,,
  \label{form7}
  \end{equation}
\begin{equation}
F(x,y,\alpha) =  x \frac{\partial\bar{\alpha}(x,y,\alpha)}{\partial x}\,.
\label{form8}
\end{equation}
Using the representation (\ref{form4}) and the definition   (\ref{form7},\ref{form8})      of the GL function 
one can find that \cite{Kr1, Kr2}
\begin{equation}
  0 \leq \psi(y, \alpha)=  \int^{\infty}_{0} \frac{t\bar{\rho}(t, \alpha,y)}{(1 +t)^2}  dt \leq
    \int^{\infty}_{0} \frac{\bar{\rho}(t, \alpha, y)}{1 +t}  dt  = \alpha \,.
 \label{form9}
\end{equation}

Let us define
\begin{equation}
  \frac{\bar{\alpha}_{2 +n}}{(q^2)^{1 + n}} \equiv (\frac{d}{dq^2})^n (\frac{\bar{\alpha}_2}{q^2}) \,,
\label{form11}
\end{equation}  
where $ \frac{{\bar{\alpha}}_2(\frac{q^2}{\mu^2}, y, \alpha)}{q^2}  =
\frac{d}{dq^2}\bar{\alpha}(\frac{q^2}{\mu^2}, \frac{m^2}{\mu^2},\alpha)$.
Using the KL representation (\ref{form1}) one can find that
\begin{equation}
  \frac{\bar{\alpha}_{2 +n}}{(q^2)^{1 + n}}= (-1)^n(n +1)! \int_0^{\infty} \frac{\rho_1(t, y, \alpha)}{(t + q^2)^{2 +n}}dt \,.
\label{form12}
\end{equation}
where $q^2 = - k^2 \geq 0$ and $\rho_1(t, y, \alpha) = \alpha_0 t \rho(t, \alpha_0, m)   \geq 0$.
Since we are  interested in the ultraviolet asymptotics we  neglect the mass $m$, i.e. we put  $y = 0$
in our formulae.\footnote{In the perturbation theory the massless limit $y \rightarrow 0$ for
  the GL function exists in
  each order of the  perturbation theory.}
The renormalization group equation for the invariant charge $\bar{\alpha}_{2+n}(\frac{q^2}{\mu^2}, \alpha) \equiv
\bar{\alpha}_{2+n}(\frac{q^2}{\mu^2},y = 0, \alpha) $ has the
form
\begin{equation}
  q^2\frac{d\bar{\alpha}_{2+n}}{dq^2}  =\beta(\bar{\alpha}) \frac{d\bar{\alpha}_{2+n}}{d\bar{\alpha}} \,.
  \label{form13}
  \end{equation}
As a consequence of the renormalization group equation (\ref{form13}) we find that
$\bar{\alpha}_{2+n}(\frac{q^2}{\mu^2}, \alpha) = \bar{\alpha}_{2+n}(1, \bar{\alpha}(\frac{q^2}{\mu^2}, \alpha))$.
Here   $q^2\frac{d\bar{\alpha}(\frac{q^2}{\mu^2}, \alpha)}{dq^2} = \beta(\bar{\alpha}(\frac{q^2}{\mu^2}, \alpha))$
 and  $\bar{\alpha}(1, \alpha)  = \alpha $.

Using the definition (\ref{form11}) of the $\bar{\alpha}_{2+n} $ and the renormalization group equation (\ref{form13})
we obtain
\begin{equation}
  \bar{\alpha}_{2+n+1}(\bar{\alpha}) = -(1+n)\bar{\alpha}_{2+n}(\bar{\alpha}) + \frac {d\bar{\alpha}_{2 + n}(\bar{\alpha})}{d\bar{\alpha}}
    \beta(\bar{\alpha}) \,.
  \label{form14}
  \end{equation}
   As a consequence of the relation (\ref{form14}) we find in particular  that
\begin{equation}
  \bar{\alpha}_{3}(\bar{\alpha}) = -\bar{\alpha}_{2}(\bar{\alpha}) + \frac {d\bar{\alpha}_{2 }(\bar{\alpha})}{d\bar{\alpha}}
    \beta(\bar{\alpha}) \,,
  \label{form15}
  \end{equation}
\begin{equation}
  \bar{\alpha}_{4}(\bar{\alpha}) = -2\bar{\alpha}_{3}(\bar{\alpha}) + \frac {d\bar{\alpha}_{3}(\bar{\alpha})}{d\bar{\alpha}} 
    \beta(\bar{\alpha}) \,.
  \label{form16}
  \end{equation}
Using the KL representation (\ref{form12}) for   $\frac{\bar{\alpha}_{2 +n}}{(q^2)^{1 + n}}$ and the non negativity 
 of the  spectral density  $\rho_1(t, \alpha) \geq 0$     we deribe 
the inequality
\begin{equation}
  0 \leq (-1)^{n+1} \bar{\alpha}_{2 + n + 1}(\bar{\alpha}) \leq (-1)^n  (n +2) \bar{\alpha}_{2 + n }(\bar{\alpha}) \,.
  \label{form17}
\end{equation}
As a consequence of the formula (\ref{form14}) the inequality (\ref{form17}) takes the form
\begin{equation}
  (-1)^{n+1} (1+n) \bar{\alpha}_{2 +n}(\bar{\alpha})
   \leq (-1)^{n+1} \frac{d\bar{\alpha}_{2+n}(\bar{\alpha})}{d\bar\alpha}
  \beta(\bar{\alpha})
   \leq    (-1)^n \bar{\alpha}_{2+n}(\bar{\alpha}) \,.
  \label{form17a}
\end{equation}
For $n = 0$ and $n =1$ the inequality (\ref{form17}) has  the form
\begin{equation}
  - \bar{\alpha}_{2}(\bar{\alpha})
   \leq - \frac{d \bar{\alpha}_{2}(\bar{\alpha})}{d\bar\alpha}
  \beta(\bar{\alpha})
   \leq \bar{\alpha}_{2}(\bar{\alpha})  \,,
  \label{form18}
\end{equation}
\begin{equation}
-2\bar{\alpha}_2(\bar{\alpha}) + 3 \frac{d\bar{\alpha}_2(\bar{\alpha})}{d\bar{\alpha}}\beta(\bar{\alpha})
\leq \frac{d}{d\bar{\alpha}}
[\frac{d\bar{\alpha}_2(\bar{\alpha})}{d\bar{\alpha}}
\beta(\bar{\alpha})]\beta({\bar{\alpha}})
   \leq \bar{\alpha}_2(\bar{\alpha}) \,.
      \label{form20}
\end{equation}


\subsection{Inequalities for the GL function in QED}

In QED in the MOM  renormalization scheme
the radiative corrections to the photon propagator at $q^2 = \mu^2$  are equal to zero. As a consequence we find that 
$\alpha_2(\bar{\alpha}) = \psi({\bar{\alpha}})$ and $\beta(\bar{\alpha}) = \psi(\bar{\alpha})$. Here   $\psi(\bar{\alpha})$
is  the GL function.
The inequalities (\ref{form18}) and (\ref{form20}) take the form
\begin{equation}
  |\frac{d\psi(\bar{\alpha})}{d\bar{\alpha}}|  \leq 1 \,,
  \label{form21}
\end{equation}
\begin{equation}
-2   +    3
 \frac{d\psi(\bar{\alpha})}{d\bar{\alpha}}
 \leq \frac{d}{d\bar{\alpha}}[ \frac{d\psi(\bar{\alpha})}{d\bar{\alpha}}
 \psi(\bar{\alpha})] \leq 1  \,.
\label{form22}
\end{equation}
Note that  in the  MOM scheme we define the coupling constant $\alpha$ as
$\alpha = \bar{\alpha}(1, \alpha)$ with $\bar{\alpha}$ defined  by the equation (\ref{form4}) to be proportional
to the transverse part of the photon propagator.  In general renormalization scheme the renormalization condition (\ref{form3})
  takes the form
\begin{equation}
  \bar{\alpha}(1, \alpha) \equiv c(\alpha) = \alpha + \sum_{k=2}^{\infty}c_k\alpha^k \,.
    \label{form23}
\end{equation}
The renormalizarion group equation for the invariant  charge $\bar{\alpha}(\frac{q^2}{\mu^2}, \alpha)$ in
arbitrary renormalization scheme has the form
\begin{equation}
  (\mu^2\frac{d}{d\mu^2} + \beta(\alpha)\frac{d}{d\alpha})\bar{\alpha}(x, \alpha) = 0\,,
\label{form24}
\end{equation}
The solution of the renormalization group equation (\ref{form24})
is 
\begin{equation}
\bar{\alpha}(x, \alpha) = \bar{\alpha}(1, \bar{\alpha}'(x,\alpha)) = c(\bar{\alpha}^{'}(x,\alpha)) \,,
\label{form27}
\end{equation}
where
\begin{equation}
x\frac{\bar{\alpha'}(x,\alpha)}{dx} = \beta(\bar{\alpha'}) \,.    
\label{form25}
\end{equation}
\begin{equation}
  \bar{\alpha'}(1,\alpha) = \alpha \,.
\label{form26}
\end{equation}    
The principal difference between the MOM scheme and arbitrary scheme is that in the MOM scheme
radiative corrections to the photon propagator dissapear at $q^2 = \mu^2$ 
 and  $c(\alpha) = \alpha $. Therefore the knowledge of the  GL function allows to restore completely the
dependence of the photon propagator on the $q^2$ while in arbitrary renormalization scheme we have to know
two functions $c(\alpha)$ and
$\beta(\alpha)$.

\subsection{Bounds for the GL function in the Wess-Zumino model}

In the derivation of the bounds for the GL function in QED we used the fact that in QED 
the invariant charge is proportional to the photon propagator for which the KL representation is valid.
This fact in general is not valid for arbitrary renormalizable field theory. However there are exceptions.
For instance, in the Wess-Zumino supersymmetric model \cite{WZ1, WZ2, WZ3} the invariant charge is
proportional to the scalar propagator. The Wess-Zumino model describes the interaction of
the scalar and Majorana fields. In the superspace the Lagrangian of the model has the form
\begin{equation}
  L = \int \phi^*(x,\theta, \bar{\theta})\phi(x,\theta, \bar{\theta})   d^2\theta d^2\bar{\theta}     +
W   + W^* \,,
\label{WZ1}
\end{equation}
\begin{equation}
 W =  \int [\frac{g}{3!}\phi^3(x, \theta) +m\frac{\phi^2(x,\theta)}{2}]  d^2\theta         \,.
\label{WZ2}
\end{equation}
Here $\phi(x, \theta) = \phi(x) + \sqrt{2}\psi(x)\theta + \theta \theta F(x)$ is chiral scalar superfield.
For the Wess-Zumino model the superpotential $W$ is not renormalized. As a consequence the
superfield $g^{1/3}\phi(x, \theta)$  is renormalization group invariant and the GL function is proportional
\cite{WZ1, WZ2, WZ3}  to the anomalous dimension $\gamma(g)$ of the scalar field $\phi(x)$, namely $\beta(g) = 3g\gamma(g)$.
It means that in analogy with QED we can define the invariant charge to be proportional to the scalar
propagator, namely
\begin{equation}
  \bar{\alpha}_{WZ}(\frac{q^2}{\mu^2}, \frac{m^2}{\mu^2}, \alpha) = -\alpha q^2 D_{\phi\phi}(-q^2, m^2, \mu^2,
\alpha) \,,
\label{WZ3}
\end{equation}
\begin{equation}
D_{\phi\phi}(p^2, m^2, \mu^2, \alpha) = i \int d^4x \exp(ipx)<0|T(\phi(x)\phi^{*}(x))|0> \,,
\label{WZ4}
\end{equation}
where $\alpha = g^{2/3} $ and $q^2 = - p^2$.
For the scalar propagator (\ref{WZ4}) the KL representation has the form
\begin{equation}
  D_{\phi\phi}(p^2, m^2, \mu^2, \alpha) = \frac{1}{p^2 - m^2_{pole}} + \int_{4m^2_{pole}}^{\infty} \frac{\rho_{\phi\phi^*}(t, m^2, \mu^, \alpha)}{p^2 -t - i\epsilon} \,.
    \label{WZ3a}
 \end{equation}

For simplicity we  consider the massless case $m = 0$. In full analogy with QED 
the renormalization group equation for the invariant charge (\ref{WZ3}) has the form \cite{bog1, Pet, Gell}
\begin{equation}
  x \frac{\partial\bar{\alpha_{WZ}}(x,\alpha)}{\partial x} = \psi_{WZ}( \bar{\alpha}) \,.
\label{WZ5}
\end{equation}
Also we use the renormalization scheme with
\begin{equation}
  \bar{\alpha}_{WZ}(x=1, \alpha) =  \alpha \,.
  \label{WZ6}
  \end{equation}
As a consequence of the KL representation (\ref{WZ3a}) for the scalar propagator and the renormalization
condition (\ref{WZ6})  we find that the inequalities
  (\ref{GL1}, \ref{GL2}, \ref{form17a}) are also valid for the GL function $ \psi_{WZ}(\alpha)$ in the Wess-Zumino model.
  It should be noted that in the Wess-Zumino model  the invariant charge $\bar{g}(\frac{q^2}{\mu^2}, g)$ could be defined also
  as the product of
  the  three point vertex and the propagators $\bar{g}(\frac{q^2}{\mu^2}, g) =
  \Gamma_3(\frac{q^2}{\mu^2}, g)(q^2D(q^2, \mu^2 ,g))^{3/2} $.
  For such definition of the invariant charge the beta function is
  \begin{equation}
    \beta(g) = \beta_1 g^3 + \beta_2 g^5 + ...\beta_n g^{2n+1} + ...  \,.
    \label{WZ7}
  \end{equation}
  As is well known  in renormalizable  models with single coupling constant
  one-loop and two-loop contributions  to the beta-function don't depend on the renormalization scheme \cite{VLADIMIROV}.
  As a consequence of this fact 
  we find that
  \begin{equation}
    \psi_{WZ}(\alpha) = \frac{2}{3}(\beta_1 \alpha^4 + \beta_2\alpha^7 + ...\beta^{'}_n\alpha^{3n+1} + ... ) \,.
    \label{WZ8}
  \end{equation}
  For instance, in one-loop approximation $\beta_1 = \frac{k_1}{16\pi^2}$, $ k_1 = \frac{3}{8}$ and 
     the inequality  (\ref{GL2}) is not valid for $\alpha^3 \equiv g^2 \geq   \frac{3}{2}\frac{1}{ \sqrt{28}}\frac{16\pi^2}{k}$.

\section{Several applications}

\subsection{Bound in QED}

The GL function in QED is known up to five loops  \cite{Kat1, Kat2, Kat3}. Namely, in five-loop approximation it reads \cite{Kat3}
\begin{equation}
  \frac{\psi(\alpha)}{\pi} = 0.333*k^2 + 0.25*k^3 + 0.0499*k^4 -0.601*k^5  + 1.434*k^6\,,
  \label{form28a}
\end{equation}
where $k = \frac{\alpha}{\pi}$.
For five-loop  approximation (\ref{form28a}) the inequality (\ref{GL2})  is valid up to $\frac{\alpha_{cr}}{\pi} = 0.53 $.

\subsection{QED with N identical fermions}

As is well known the $\beta$-function in renormalizable field theory with single effective charge does not depend on the renormalization
scheme in two-loop approximation. For QED with N identical fermions in   one-loop approximation
the GL function is
\begin{equation}
  \psi(\alpha, N) = N\frac{\alpha^2}{3\pi}  \,.
\label{form28}
\end{equation}
In two-loop approximation the GL function has the form
\begin{equation}
\psi(\alpha, N)
= N[\frac{\alpha^2}{3\pi} + \frac{\alpha^3}{4\pi^2}] \,.
\label{form29}
\end{equation}
For one-loop  approximation (\ref{form28}) the inequality (\ref{GL2})  is valid up to $\frac{\alpha_{cr}}{\pi} = \frac{1.22}{N}$.
  For    $\frac{\alpha_{cr}}{\pi} = \frac{1.22}{N}$        the ratio of two-loop correction to one-loop approximation for the GL function
is equal  to $\frac{0.92}{N}$ and for
$N \geq 10$ it is less than 10 percent. One can show that higher order corrections to GL function qualitatively don't change our conclusions.
Really, the  GL function up to four loops  is  \cite{Kat1}
$$
   \psi (\alpha, N) =     N[\frac{\alpha^2}{3\pi} + \frac{\alpha^3}{4\pi^2}]   +    4\pi(\frac{\alpha}{4\pi})^4(-2 N +   (\frac{64}{3} \zeta(3) -\frac{184}{9})N^2) +
  $$
  \begin{equation}
  4\pi(\frac{\alpha}{4\pi})^5(-46N +(104 +\frac{512}{3}\zeta(3) - \frac{1280}{3}\zeta(5))N^2 + (128 -\frac{256}{3}\zeta(3))N^3) \,.
\label{form30}
  \end{equation}
  For  $\frac{\alpha_{cr}}{\pi} = \frac{1.23}{N}$ the     three and four loop    corrections  are less than 5 percent.

According to common lore we can trust one-loop approximation provided two-loop correction  is much smaller one-loop approximation.
So we can think that one-loop approximation is correct for $N \geq 10$ and $\alpha = \alpha_{cr}$. In other words we find that one-loop
approximation contradicts to the inequality  (\ref{form22}) for  $\frac{\alpha_{cr}}{\pi} = \frac{1.22}{N}$ and for $N \geq 10$ we
can trust the perturbation theory. We can interpret this result as an indication in favour of vacuum instability in QED with
$N \geq 10$  identical fermions \footnote{Note that using a $\frac{1}{N}$ expansion we immediately obtain the wrong pole for photon propagator \cite{LINDE}
  in many charged QED.}.  In perturbation theory $L \geq 2$  loop correction to the GL function for $N \gg 1 $ is proportional to
$N^{L-1}(\frac{\alpha}{\pi})^{1+L} $
and it is much smaller one-loop contribution.

\subsection{Bound on new particles contribution  in the SM extensions}

As is well known in the SM and its extensions  based on the gauge group $SU_c(3) \otimes SU_L(2) \otimes U_Y(1) $ the
GL function for  the $U_Y(1)$ subgroup in one-loop approximation 
is
\begin{equation}
  \psi(\alpha_1) =   N_{tot} \frac{\alpha^2_1}{3\pi} \,,
  \label{formula30aa}
  \end{equation}
where $N_{tot}= 5.125 + \Delta N $ and $\alpha_1 = \frac{g^2_1}{4\pi}$.  Here   $5.125$ is the contribution from quarks, leptons,
Higgs isodoublet and $\Delta N \geq 0$ is the contribution
  from new particles beyond the SM. We shall assume that the SM with the gauge group  $SU_c(3) \otimes SU_L(2) \otimes U_Y(1) $
    and possible new particles (isosinglets, isodoublets,...) is valid up to $M_{cr} = \frac{M_{PL}}{10}$ \footnote{For
       scales  $E \leq M_{cr}$  the effects from gravity are
      proportional to $\frac{1}{4\pi^2} M_{PL}^2 E^2 $  and they are not essential.}.
As a consequence of the inequality (\ref{GL2})
we find that
\begin{equation}
\bar{\alpha}_1(\frac{q^2}{\mu^2}, \alpha_1)
\leq  \frac{3\pi}{\sqrt{6}N_{tot}}  \,
\label{inequality}
\end{equation}
for $q^2 \leq M^2_{cr}$.
In the SM  $\bar{\alpha_1}(M^2_Z) = \frac{\bar{\alpha}_{em}(M^2_Z)}
{\cos^2(\theta_W)}$.  For $\bar{\alpha}_{em}(M_Z) = \frac{1}{128}$,  $\sin^2(\theta_W)  = 0.2245 $ \cite{PARTICLEDATA}
we find that $\bar{\alpha}_1(M^2_Z) = 0.010$.
The use of the solution of the renormalization group equation with the GL function (\ref{formula30aa})
and the inequality (\ref{inequality}) leads to 
\begin{equation}
  \Delta N \leq 7.1  \,.
  \label{inequality1}
\end{equation}
Here the parameter $\Delta N  = \sum_{k}Y_k^2 c_k \frac{  \ln(\frac{M^2_{cr}}{M^2_{k}})} { \ln(\frac{M^2_{cr}}{M^2_{Z}})}    $
is the contribution of
new particles with masses $M_k$ and hypercharges $Y_k$.\footnote{The parameter $c_k$ is equal to 1 for Dirac fermion,
  1/2 for Weyl fermion and 1/4 for scalar.}
The inequality (\ref{inequality1}) allows to restrict possible additional particles in the SM extension. For instance, the inequality (\ref{inequality1}) 
excludes  the existence of the fermion with the hypercharge $Y = 3$ and the mass $O(10)$~TeV and also
it excludes 
the existence of 3  additional vector like generations with the masses $O(10)$~TeV. Note that the   requirement of
the absence of Landau pole singularity for scales less than $M_{cr}$ \cite{PARISI} leads to slightly weaker bound.

\subsection{Bound on coupling constant in dark photon  model}

In dark photon model \cite{dark1} -\cite{dark3} new  massive  vector field  $A'$ (dark photon)
interacts with light dark matter \cite{light1} The interaction between the
SM particles and dark sector arises as a consequence of   nonzero kinetic mixing of photon and dark photon. In dark photon model
very important parameter is analog of electromagnetic fine structure constant $\alpha_D = \frac{e^2_D}{4\pi} $,
 where $e_D$ is dark photon charge.
To compare the predictions of dark photon model with experimental data it is necessary to know the bound on the
coupling constant $\alpha_D$. 
  Very often the variant of dark photon model with pseudo Dirac fermions  \cite{PSEUDO}  is used.
In the model with pseudo Dirac fermion besides fermion field we have to introduce scalar field with the charge $2 e_D$
and the GL function for model with dark photon field in two-loop approximation in the ultraviolet region is \cite{LANDAUPOLE}
\begin{equation}
  \psi(\alpha_D) = \frac{2\alpha^2_D}{3\pi} + \frac{5 \alpha^3_D}{4\pi^2} \,.
  \label{darkphoton1}
\end{equation}
We shall require that two-loop approximation (\ref{darkphoton1}) does not contradict to the inequality (\ref{GL2}) for
the scales up to $M_{cr} = \frac{M_{PL}}{10}= 1.2 \cdot 10^{18}~GeV$. For $M_{A'} = 1~GeV$ we find that
$\alpha_D \equiv \bar{\alpha}_D(M^2_{A'}) \leq 0.046 $.  Note that
from the requirement of the absence of Landau pole singularity for the scales $\leq M_{cr}$ \cite{LANDAUPOLE} slightly weaker  bound 
  $\alpha_D \leq 0.049 $ has been obtained.

\subsection{Bound on coupling constant in  dark matter model with $(B-L)$ vector messenger}

In this subsection we apply the inequality (\ref{GL2})  for constraining the  dark matter models with $B-L$
\cite{B-L.1, B-L.2, B-L.3, B-L.4}  vector messenger.
We assume that the dark matter is described by the fermion field $\psi_D$ with
a mass $m_D$ and coupling constant $g_D$ with $Z'$ boson.
The interaction of the $Z'$ -boson with quarks and leptons of the SM and dark matter $\psi_D$ has the form
\begin{equation}
L_{int} =  g_{B-L} (\sum_{quarks}\frac{1}{3}\bar{q}\gamma^{\mu}q - \sum_{leptons}\bar{l}\gamma^{\mu}l)Z'_{\mu} +g_D\bar{\psi}\gamma^{\mu}\psi Z_{\mu}' \,.
  \label{form31a}
\end{equation}
The underground experiments \cite{PARTICLEDATA} look for dark matter by the search for the reaction of elastic nucleon dark matter
scattering. For dark matter with the mass of dark matter particles $m_D \sim O(1)TeV$ the experimental bound on
the elastic nucleon dark matter cross section  \cite{PARTICLEDATA, XENON}
\begin{equation}
  \sigma_{el} \leq 10^{-9} \kappa_{el}~pb \,,
  \label{form31b}
\end{equation}
where $\kappa_{el} = O(1)$.
In considered model (\ref{form31a})       the elastic $DM ~+~N \rightarrow DM ~+~N$ 
nucleon dark particle  cross section is \cite{ELCROS}
\begin{equation}
  \sigma_{el} = \frac{16 \pi \alpha_{B-L}   \alpha_D m_p^2} {m^4_{Z'}} \,,
  \label{form31c}
  \end{equation}
  where $\alpha_{B-L} = \frac{g^2_{B-L}}{4\pi}$, $\alpha_{D} = \frac{g^2_{D}}{4\pi}$ and $m_p$ is the proton mass.
  Additional standard assumption is that in  the early Universe dark matter was in the
    thermodynamic equilibrium with the SM matter
  and during the Universe expansion at  some temperature dark matter  decouples. From the condition that the
  dark matter density at present epoch is  $\frac{\rho_D}{\rho_{cr}}   \approx 0.23 $ one can find that
  the annihilation cross section of $DM ~+~\bar{D M} \rightarrow SM ~particles $ is \cite{PARTICLEDATA, KOLB}
\begin{equation}
  <\sigma_{an}v> =\kappa_{an} \cdot pb\cdot c    \,, 
\label{form31e}
\end{equation}
    where $\kappa_{an} = O(1)) $  and $c$ is the velocity of the light.
  The annihilation cross section for the  model   (\ref{form31a}) is \cite{dark3}
  \begin{equation}
    \sigma_{an}v = \frac{16 \pi c_{an}  \alpha_{B-L}   \alpha_D m_{DM}^2} {(m^2_{Z'} - 4 m^2_{DM})^2 } =
    \frac{16 \pi c_{an}  \alpha_{B-L}   \alpha_D k_{DM} } { m^2_{Z'}  }  \,,
    \label{form31d}
  \end{equation}
  where $c_{an} = 9$ and $k_{DM}^{-1} = \frac{(m^2_{Z'} - 4 m^2_{DM})^2 }{m^2_{Z'}m^2_{DM}}$. For often used relation $m_{Z'} = 3m_{DM} $
  we find  $k_{DM} =    9/25$. We also assume that there is no fine tuning between $m_{Z'}$ and $m_{DM}$,
  namely we assume that $|2 m_{DM} - m_{Z'}| \geq 0.2 m_{DM} $.
  This assumption means that $k_{DM} < 7 $.
  As a consequence of the formulae (\ref{form31b}, \ref{form31c}, \ref{form31d}, \ref{form31e}) one can find that
  \begin{equation}
    \frac{\sigma_{el}}{<\sigma_{an}v>} =
  \frac{m^2_p}{m^2_{Z'}}\cdot \frac{1}{c_{an}k_{DM}}
       \leq \frac{\kappa_{el}}{\kappa_{an}c }  \cdot 10^{-9}    \,,
      \label{form31f}
      \end{equation}
  \begin{equation}
16\pi\alpha_D\alpha_{B-L} \geq 2.3 (\frac{1}{c_{an}k_{DM}})^2 \cdot \frac{\kappa_{an}^2}{\kappa_{el}}\,. 
\label{form31ff}
  \end{equation}
  
  We shall assume that for $(B-L)$ model the perturbation theory is valid for the scales  
  up to  $M_{cr} = \frac{M_{PL}}{10}$.
The GL function for $U_{B-L}(1)$ gauge group in one-loop aprroximation is $\psi(\alpha_{B-L}, \alpha_D) = \frac{1}{3\pi}(8\alpha^2_{B-L} + \alpha_D^2)$.
  As a consequence  we  find that
  in one-loop approximation\footnote{An account of two-loop correction leads to less than  four percent change in the inequality
    (\ref{form31g})}
    \begin{equation}
16\pi\alpha_{B-L}\alpha_{D} \equiv  16\pi\alpha_{B-L}(M_{Z'})\alpha_{D}(M_{Z'}) \leq 0.028.
 \label{form31g}
    \end{equation}
    The bound (\ref{form31g}) contradicts to the bound (\ref{form31f}) at $m_{Z'} = 3 m_{DM}$
    for the uncertainties
    $\frac{\kappa_{el}^2}{\kappa_{an}} \geq 0.11$.
    So we have found that the bound (\ref{form31ff}) at $m_{Z'} = 3 m_{DM}$  contradicts
    to the bound (\ref{form31g}) derived in the assumption that the perturbation theory for $(B - L$ model is valid for
    the scales up  to $10^{18}~GeV$.

\section{Conclusions}

In this paper we derived new inequalities for the GL function in QED.
The main idea of the derivation is the use of the fact that the
invariant charge and the GL function in QED are determined by the transverse part of the photon
propagator. The KL representation with non negative spectral  density is valid
for transverse part of the photon propagator that is crusial ingredient for the derivation of the
bounds on the GL function in QED.
Also we  derived analogous bound for the
supersymmetric Wess-Zumino model where the invariant charge is proportional to
the scalar propagator.
As practical applications we have considered models based on the abelian gauge group $U(1)$.
In the assumption that the perturbation theory is valid up to some scale $M_{cr}$ we have determined the range of the
$U(1)$ coupling constant and obtained the bounds on the effective coupling constant at low scale.
Bounds on the effective coupling constant at low scale allow
    to
restrict  free parameters of  dark photon model and  dark matter model based on the vector $(B-L)$ messenger.

I  am  indebted to  A.L. Kataev for useful conversations and for pointing out to me some references. Also I thank 
 the collaborators of the INR theoretical department   
for discussions and critical comments.

\end{document}